\documentclass[11pt]{article}
\usepackage{moreverb}
\usepackage[superscript,biblabel]{cite}
\usepackage[utf8]{inputenc}
\usepackage{amsfonts}
\usepackage{rotating}
\usepackage{tikz}
\usepackage{multirow}

\usepackage{amsmath}
\usepackage[margin=1in]{geometry}
\usepackage{multicol}
\usepackage{graphicx}
\usepackage{multicol}
\usepackage{float}
\usepackage{verbatim}
\newcommand{\RNum}[1]{\uppercase\expandafter{\romannumeral #1\relax}}
\begin{document}

\title{A time-dependent Poisson-Gamma model for recruitment forecasting in multicenter studies}

\author{Armando Turchetta\thanks{Department of Epidemiology, Biostatistics and Occupational Health, McGill University} ,
Nicolas Savy\thanks{Toulouse Mathematics Institute, University of Toulouse III} ,
David A. Stephens\thanks{Department of Mathematics and Statistics, McGill University } ,
Erica E.M. Moodie\footnotemark[1] ,
Marina B. Klein\thanks{Department of Medicine, Division of Infectious Diseases/Chronic Viral Illness Service, McGill University Health Center}
}

\date{}

\maketitle

\begin{abstract}
Forecasting recruitments is a key component of the monitoring phase of multicenter studies. One of the most popular techniques in this field is the Poisson-Gamma recruitment model, a Bayesian technique built on a doubly stochastic Poisson process. This approach is based on the modeling of enrollments as a Poisson process where the recruitment rates are assumed to be constant over time and to follow a common Gamma prior distribution. However, the constant-rate assumption is a restrictive limitation that is rarely appropriate for applications in real studies. In this paper, we illustrate a flexible generalization of this methodology which allows the enrollment rates to vary over time by modeling them through B-splines. We show the suitability of this approach for a wide range of recruitment behaviors in a simulation study and by estimating the recruitment progression of the Canadian Co-infection Cohort (CCC).\end{abstract}

\section{Introduction}
The recruitment time estimation is an important topic of interest in the planning and monitoring phase of multicenter studies with several practical implications. Yet, deterministic models mainly based on the study investigators' recruitment estimates are still used \cite{as4, sav2}. This can lead to (1) ignoring important sources of variability and (2) overestimation of recruitment rates. The latter consequence is due to a phenomenon commonly known as the `Lasagna Law',\cite{lasagna,rerct2} according to which medical investigators tend to be overly optimistic regarding the number of participants who meet the inclusion criteria and are willing to enroll in the study. These factors often lead to an underestimation of recruitment times and thus to difficulties in achieving the targeted sample size. For example, Walters \& al.~\cite{recrct} reviewed 151 RCTs conducted in the United Kingdom, concluding that 44\% of the trials did not achieve the final recruitment target. A similar estimate was found by van der Wouden \& al.~\cite{rerct2} analyzing 78 primary care research studies conducted in the Netherlands, who also noted that 51\% of the studies had to extend the fieldwork period, and of these, 79\% needed an extension longer than 50\% of the originally planned study length. \\
A Bayesian approach built on a doubly stochastic Poisson process, known as the Poisson-Gamma model, was introduced to address the lack of a strong and consistent statistical methodology in this field in a series of papers by Anisimov \& al.~\cite{as3,as5,as4,as1,as2} This approach is based on modeling participants' arrival to recruitment centers as a Poisson process where the recruitment rates are assumed to be constant over time and to originate from a common Gamma distribution whose parameters are estimated from the ongoing trial in an empirical Bayes fashion. The model has been validated using several real trials' recruitment data and found to have good performances when the number of centers involved in the study is sufficiently high ( $>$20). The Poisson-Gamma recruitment model has been further extended in numerous directions: Mijoule \& al.~\cite{sav1} considered the use of the Pareto distribution in place of the Gamma distribution, Bakhshi \& al.~\cite{bak} and Minois \& al.~\cite{sav2} suggested methods to exploit historical data from previous trials to estimate recruitment predictions before the start of the trial, Minois \& al.~\cite{sav3} accounted for breaks in the recruitment process, and Anisimov \& al.~\cite{sav4} augmented the model to account for drop-outs.  The use of the Poisson process to capture the recruitment progression in clinical trials has been largely accepted in the literature and has deep roots. To the best of our knowledge, this framework was first introduced in Lee \cite{lee}, however, for a systematic review of early recruitment models for multicenter clinical trials we refer the reader to Barnard \& al.~\cite{review}
More recently, Gajewski \& al.~\cite{gaj} modeled the waiting times between participants as exponential random variables (which is equivalent to a Poisson process) incorporating subjective knowledge on the recruitment process through an informative prior distribution, and Jiang \& al.~\cite{jiang} further elaborated this model by mitigating overly-optimistic investigators' assessments via adaptive priors. \\
All the aforementioned methods assume that the recruitment intensity remains constant over time. However, this assumption is seldom met in practice. A number of authors have proposed methods to mitigate this pitfall. Tang \& al.~\cite{tang} proposed a discrete-time Poisson process-based method using a piece-wise linear function to model the accrual rate, where changepoints can be either fixed or estimated. However, this model was mostly tailored to their specific clinical trial example.
A more general approach was introduced in Zhang \cite{zhang}, where the authors proposed a Bayesian method that relaxes the assumption of constant accrual rate via a non-homogeneous Poisson process where the overall underlying time-dependent accrual rate is modeled through cubic B-splines. This approach was further extended by Deng \& al.~\cite{deng} in order to accommodate staggered initiation times and differences in accrual rates across regions. More recently, building on the standard Poisson-Gamma model, Lan \& al.~\cite{lan} proposed a model where recruitment rates are assumed to be constant up to a certain point and then decay over time as a negative exponential.
Urbas \& al.~\cite{urban} further expanded this idea allowing for the detection of time-inhomogeneity via a testing procedure and considering a wider range of parametric curves to model the decaying recruitment evolution over time. All these methods performed well in the recruitment scenarios they were built to target.\\
In this paper, we present a novel extension of the Poisson-Gamma model which relaxes the constant-rate assumption. We model the recruitment process as a non-homogeneous Poisson process where the rates originate from a Gamma distribution whose mean parameter depends on time. Specifically,  we assume that, at some unknown point in time, the recruitment progression will reach a plateau, and the non-constant section is modeled via B-splines; note that this approach is applicable to any multicenter study, whether a randomized trial or a cohort study. We outline the details of this methodology in Section 2. In Section 3, we assess in a simulation study the performance of the proposed approach in relation to the standard Poisson-Gamma model for a variety of recruitment behaviors. In Section 4, we apply our method to predict enrollments in the Canadian Co-infection Cohort (CCC). Section 5 concludes.

\section{Methodology}\label{met}

Let $n_i(t)$ be the number of participants enrolled up to time $t$ in center $i$, $C$ the number of centers,  $N(t)=\sum_{i=1}^{C}n_{i}(t)$ the total enrollments at time $t$, and $u_i$ the center-specific initiation time.

\subsection{The Poisson-Gamma model}
The Poisson-Gamma (PG) model \cite{as3,as5,as4,as1,as2} assumes that participants arrive at each center according to a Poisson process with rate $\lambda_i(t)=I_{\{t>u_i\}}\lambda_i$.  According to this model, the $\lambda_i$'s are viewed as a sample from a Gamma distribution with parameters ($\alpha, \beta$) and probability density function
\[p(x|\alpha,\beta)=\frac{\beta^{\alpha}}{\Gamma(\alpha)}e^{-\beta x}x^{\alpha-1}.\]
Hence, $n_i(t)$ is a Poisson process with cumulative rate
\[
    \Lambda_i(t)=I_{\{t>u_i\}}(t-u_i)\lambda_i
\]
and, due to the superposition theorem of independent Poisson point processes, $N(t)$ is a Poisson process with cumulative rate
\[
    \Lambda(t)=\sum_{i=1}^{C}\Lambda_i(t)=\sum_{i=1}^{C}I_{\{t>u_i\}}(t-u_i)\lambda_i.
\]
Consequently, assuming that we have reached a point in time $t_{int}$ where all the $C$ centers have started the enrollment phase and indicating with $k_i$ the cumulative number of recruited participants in each center and with $\tau_i=t_{int}-u_i$ the time elapsed between the centers' initiation and $t_{int}$, the posterior density of each rate $\lambda_i$ is a Gamma distribution with parameters $\alpha+k_i$ and $\beta + \tau_i$.\\
The hyperparameters $\alpha$ and $\beta$ can be estimated in an empirical Bayes fashion using the enrollment data collected up to $t_{int}$. Since the conditional distribution of $k_i$ is Poisson with rate parameter $\tau_i\lambda_i$,  its marginal distribution is 
\[
    k_i\sim \textrm{NB}\left( \alpha,\frac{\tau_i}{\beta}\right)
\]
where NB indicates a negative binomial random variable.
Setting $m=\alpha/\beta$ and assuming that participants are recruited independently in the $C$ centers, $\alpha$ and $m$  can be estimated by maximizing the following likelihood:
\[
    \mathcal{L}(\alpha, \beta, \pmb{k},\pmb{\tau})=\prod_{i=1}^{C}{\alpha + k_i -1 \choose \alpha -1}\left(\frac{m \tau_i}{\alpha + m \tau_i}\right)^{k_i}\left(\frac{\alpha}{\alpha + m \tau_i}\right)^\alpha .
\]

It follows that cumulative rate $\Lambda$ is distributed as a sum of Gamma random variables, i.e.
\[
    \widetilde\Lambda=\sum_{i=1}^{C}\textrm{Gamma}(\widehat\alpha + k_i,\, \widehat\beta + \tau_i).
\]
Hence, the (additional) total number of enrollments $N^a(T)$ at time $T$ is $N^a(T)=\textrm{Poisson}(\widetilde\Lambda T)$. However, the constant-rate assumption is a restrictive limitation that is rarely appropriate for real applications in clinical studies. In the next section, we generalize the Poisson-Gamma model to the time dependence of the recruitment rates.


\subsection{The time-dependent Poisson-Gamma model}

\subsubsection{Overview}
 We assume that participants arrive at each center according to a non-homogeneous Poisson process with rate $\lambda_i(t)$, where the rates are viewed as originating from a Gamma distribution whose mean parameter depends on time.  Discretizing time in a pre-specified time unit, say days, let us indicate with $t_1$ the first day of the recruitment process, with $t_{int}$ the interim time when the predictions on future enrollments are estimated, and with $T$ the final point in time for which we estimate predictions. We assume that the recruitment intensity will plateau at an unknown point in time $t_p$. The main idea behind this methodology is to still employ the Poisson-Gamma recruitment model, but only using the recruitment data collected after the rates plateau. Each center initiates its recruitment process at time $u_i$, hence $\pmb{t^i}=\{t_1, t_2, \dots , t-u_i\}$ indicates the days elapsed between the center's initiation time and $t$.
The model can be expressed as follows:
\begin{align*}
  &  N(t)\sim \text{Poisson}[\Lambda(t)],\\
   &     \Lambda(t)=\sum_{i=1}^{C}\Lambda_i(t)=
    \sum_{i=1}^{C}\int_{t_1}^{t-u_i}\lambda_i(s)ds=
    \sum_{i=1}^{C}\sum_{s=t_1}^{t-u_i}  \lambda_i(s),\\
    & \lambda_i(t)  \begin{cases}
  \sim \textrm{Gamma}\left(\alpha, \frac{\alpha}{m(\pmb{\phi},t)}\right) \qquad t\leq t_p \\
     = \lambda_i(t_p) \qquad\qquad\qquad\qquad\,\, t> t_p
    \end{cases} 
\end{align*}
where $\pmb{\phi}$ is a set of parameters that needs to be estimated. Note that for $t_p=1$ this model corresponds to the standard Poisson-Gamma model. We model the non-constant section of the rates' evolution over time through B-splines.\cite{spln} Specifically, $m(\pmb{\phi},t)$ is modelled as follows:
\[
\log(m(\pmb{\phi},t))=
    \begin{cases}
     \sum_{k=1}^d \eta_k\gamma_k(t) & t<t_p\\
      \eta_d & t\geq t_p,\\
    \end{cases} 
\]
where $\gamma_1(t),\ldots, \gamma_d(t)$ represent the basis functions.
 
\subsubsection{Estimation}
\begin{figure}[H]
  \centering
\includegraphics[width=0.75\linewidth]{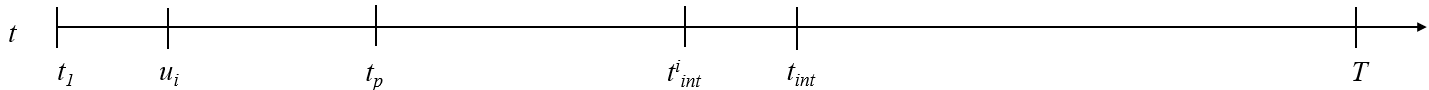}
\caption{Timeline: $t_1$ is the start of the recruitment process, $u_i$ the initiation time for center $i$, $t_p$ the point in time where the plateau begins, $t_{int}$ the interim time, $t^{i}_{int}$ the time center $i$ has been recruiting until $t_{int}$, and $T$ the final point in time for which predictions are estimated. }
\label{fig:timeline}
\end{figure}

Let $t^{i}_{int}$ represent the number of days each center has been recruiting until the interim time, i.e. $t^{i}_{int}=t_{int}-u_i+1$. Figure \ref{fig:timeline} summarizes the notation of the timeline.
We separate the $C$ centers into two groups, the first group containing the centers that have already passed the plateau point and the second group containing the ones that have not.   Sorting the centers by initiation time, we indicate with $C^*$ the number of centers that belong to the first group, that is $C^*=\text{card}\{i:t^{i}_{int}\geq {t}_p\}$.
Assuming that at time $t_{int}$ all the centers have started the enrollment process, that is $u_i<t_{int} \,\,\, \forall i$, the conditional distribution of the number of recruited participants in each center $N_i(t_{int})$ is Poisson with rate parameter $\Lambda_i(t_{int})=\sum_{s=t_1}^{t^i_{int}}\lambda_i(s)$, hence the conditional joint distribution of daily enrollments is
\[\begin{split}
\Pr\left[n_i(t_1),...,n_i(t^i_{int})|\lambda_i(t_1),...,\lambda_i(t^i_{int})\right]&\propto e^{-\sum_{s=t_1}^{t^i_{int}}\lambda_i(s)}\prod_{s=t_1}^{t^i_{int}} \lambda_i(s)^{n_i(s)}\\
&=\left[e^{-(t^i_{int}-t_p+1)\lambda_i(t_p)}\lambda_i(t_p)^{\sum_{s=t_p}^{t^i_{int}}n_i(s)}\right]^{1_{\{i\leq C^*\}}}\\
&\quad e^{-\sum_{s=t_1}^{(t_{p}-1) \wedge (t^i_{int})}\lambda_i(s)}\prod_{s=t_1}^{(t_{p}-1) \wedge t^i_{int}} \lambda_i(s)^{n_i(s)},
\end{split}\]
where $1_{\{i\leq C^*\}}$ is the indicator function which denotes the centers whose recruitment process has already passed the plateau point and $(t_{p}-1) \wedge t^i_{int}=\min\{(t_{p}-1),t^i_{int}\}$.
Setting $N_i^*(t)=\sum_{s=t_p}^{t}n_i(s)$, the marginal predictive distribution is
\[\begin{split}
\Pr\left[n_i(t_1),...,n_i(t^i_{int})\right]&=\int_{0}^\infty...\int_{0}^\infty \Pr\left[n_i(t_1),...,n_i(t^i_{int})|\lambda_i(t_1),...,\lambda_i(t^i_{int})\right]\pi(\lambda_i(t_1),...,\lambda_i(t^i_{int}))\prod_{s=t_1}^{t^i_{int}}d\lambda_i(s)
\\
&\propto
\left[\frac{\Gamma(\alpha+N_i^*(t^i_{int}))}{\Gamma(\alpha)}
\left(\frac{m(\pmb{\phi},t_{p}) }{\alpha + m(\pmb{\phi},t_{p})  }\right)^{N_i^*(t^i_{int})}\left(\frac{\alpha}{\alpha + m(\pmb{\phi},t_{p})  }\right)^{(t^i_{int}-t_p+1)\alpha}\right]^{1_{\{i\leq C^*\}}}\\
&\qquad\quad\prod_{s={t}_{1}}^{(t_{p}-1) \wedge t^i_{int} }
\frac{\Gamma(\alpha+n_i(t_j))}{\Gamma(\alpha)}
\left(\frac{m(\pmb{\phi},s) }{\alpha + m(\pmb{\phi},s)  }\right)^{n_i(s)}\left(\frac{\alpha}{\alpha + m(\pmb{\phi},s)  }\right)^\alpha.
\end{split}
\]
Finally, assuming independence between the number of participants recruited in different centers, the likelihood function is
\begin{align*}
\mathcal{L}(\alpha, \pmb{\phi},\pmb{n})\propto
\prod_{i=1}^{C}\left[\frac{\Gamma(\alpha+N_i^*(t^i_{int}))}{\Gamma(\alpha)}
\left(\frac{m(\pmb{\phi},t_{p}) }{\alpha + m(\pmb{\phi},t_{p})  }\right)^{N_i^*(t^i_{int})}\left(\frac{\alpha}{\alpha + m(\pmb{\phi},t_{p})  }\right)^{(t^i_{int}-t_p+1)\alpha}\right]^{1_{\{i\leq C^*\}}}\\
\qquad\quad\prod_{s={t}_{1}}^{(t_{p}-1) \wedge t^i_{int} }
\frac{\Gamma(\alpha+n_i(t_j))}{\Gamma(\alpha)}
\left(\frac{m(\pmb{\phi},s) }{\alpha + m(\pmb{\phi},s)  }\right)^{n_i(s)}\left(\frac{\alpha}{\alpha + m(\pmb{\phi},s)  }\right)^\alpha,
\end{align*}
 where $\pmb{n}$ represents the observed enrollments from the $C$ centers at every time point. Using this likelihood, $\alpha$ and $\pmb{\phi}=\{t_p, \eta_1,\dots \eta_d\}$ can be estimated and plugged into the posterior distribution of $\lambda_i(t)$.

\subsubsection{Spline model choice}

To guarantee enough flexibility in the recruitment curves that this model is able to estimate, we proposed to model the time-dependent enrollment intensity through B-splines. However, this choice implies that decisions have to be made regarding the degree of the polynomials, the number of knots, and their placement. Since these factors affect the dimension of the parameter space, we propose to fit different models and select the one which is best-fitting using the Bayesian Information Criterion (BIC) \cite{bic}.

\subsubsection{Forecasting future enrollments}

For the centers that have already passed the plateau point, we can project the posterior distribution of $\lambda_i(t_{int})$ to predict future enrollments, so that the distribution of the additional recruited participants for center $i\in\{1,...,C^*\}$ at time $T>t_{int}$ is Poisson with rate $\Lambda_i^a(T)=(T-t_{int})\lambda_i(t_{int})$, where
 \[
\lambda_i(t_{int})\sim\text{Gamma}\left(\widehat{\alpha} +\sum_{s=\widehat{t}_{p}}^{t^i_{int}} n_i(s), \,\,\, \beta(t,\widehat{\pmb{\phi}})+t^i_{int}-\widehat{t}_p+1\right ), \qquad i=1,...,C^*,
 \]
and $\beta(t,\pmb{\phi})=\alpha/m(\pmb{\phi},t)$.
On the other hand, for the remaining centers $i\in \{C^*+1,...,C\}$, 
\begin{align*}
& \Lambda_i^a(T)=
\sum_{s=t^{i}_{int}+1}^{ \widehat{t}_{p}-1}\lambda_i(s)+(T-u_i-\widehat{t}_{p})_+\lambda_i(\widehat{t}_{p}),\\
&\lambda_i(t)\sim\text{Gamma}\left(\widehat{\alpha} , \,\,\, \beta(t,\widehat{\pmb{\phi}})\right ),
\end{align*}
where $(T-u_i-\widehat{t}_{p})_+=\text{max}(0,T-u_i-\widehat{t}_{p})$. Therefore, setting $t^i_{p}=t_p+u_i$, the distribution of future additional enrollments at time $T$, $N^a(T)$, is Poisson with overall recruitment rate
\[\Lambda^a(T)=\sum_{i=1}^{C^*}(T-t_{int})\lambda_i(t_{int}) +
\sum_{i=C^*+1}^C\sum_{s\in \{t^{i}_{int},\dots, \widehat{t}^{i}_{p}\}}\lambda_i(s)+
\sum_{i=C^*+1}^C(T-\widehat{t}^i_{p})_+\lambda_i(\widehat{t}_{p}).
\]
It follows that the expectation and variance of the distribution of future enrollments are
\begin{align*}
E[N^a(T)]&= E\{E[N^a(T)|\Lambda^a(T)]\}=E[\Lambda^a(T)],\\
Var[N^a(T)]&= Var\{E[N^a(T)|\Lambda^a(T)]\}+E\{Var[N^a(T)|\Lambda^a(T)]\}=Var[\Lambda^a(T)]+E[\Lambda^a(T)],
\end{align*}
where
\[
\begin{split}
E[\Lambda^a(T)]&=\sum_{i=1}^{C^*}(T-t_{int})E[\lambda_i(t_{int})] +
\sum_{i=C^*+1}^C\sum_{s\in \{t^{i}_{int},\dots, \widehat{t}^{i}_{p}\}}E[\lambda_i(s)]+
\sum_{i=C^*+1}^C(T-\widehat{t}^i_{p})_+E[\lambda_i(\widehat{t}_{p})]\\
&=(T-t_{int})\sum_{i=1}^{C^*}\frac{\widehat{\alpha} +\sum_{s\in \{\widehat{t}^{i}_{p},\dots, t_{int}\}} n_i(s)}{\beta(t,\widehat{\pmb{\phi}})+t_{int}-\widehat{t}^{i}_{p}} +
\sum_{i=C^*+1}^C\sum_{s\in \{t^{i}_{int},\dots, \widehat{t}^{i}_{p}\}}\frac{\widehat{\alpha} }{\beta(s,\widehat{\pmb{\phi}})}+\\
&\qquad \qquad \sum_{i=C^*+1}^C(T-\widehat{t}^i_{p})_+\frac{\widehat{\alpha} }{\beta(\widehat{t}_{p},\widehat{\pmb{\phi}})},
\end{split}\]
\[\begin{split}
Var[\Lambda^a(T)]&=\sum_{i=1}^{C^*}(T-t_{int})^2Var[\lambda_i(t_{int})] +
\sum_{i=C^*+1}^C\sum_{s\in \{t^{i}_{int},\dots, \widehat{t}^{i}_{p}\}}Var[\lambda_i(s)]+
\sum_{i=C^*+1}^C(T-\widehat{t}^i_{p})_+^2Var[\lambda_i(\widehat{t}_{p})]\\
&=(T-t_{int})^2\sum_{i=1}^{C^*}\frac{\widehat{\alpha} +\sum_{s\in \{\widehat{t}^{i}_{p},\dots, t_{int}\}} n_i(s)}{(\beta(t,\widehat{\pmb{\phi}})+t_{int}-\widehat{t}^{i}_{p})^2} +
\sum_{i=C^*+1}^C\sum_{s\in \{t^{i}_{int},\dots, \widehat{t}^{i}_{p}\}}\frac{\widehat{\alpha} }{\beta(s,\widehat{\pmb{\phi}})^2}+\\
& \qquad \qquad
\sum_{i=C^*+1}^C(T-\widehat{t}^i_{p})_+^2\frac{\widehat{\alpha} }{\beta(\widehat{t}_{p},\widehat{\pmb{\phi}})^2}.
\end{split}
\]
Using these quantities, we can construct credible intervals (CrIs) for future enrollments using the Normal approximation. If the quantity of interest is the remaining time to reach the target number of recruitments, one can compute the expectations and credible intervals for future enrollments for a grid of values of $T$ and then invert them to obtain a point estimate and credible interval of the remaining time to recruit the target number of participants.

\section{Simulation study} \label{sim}

In this section, we analyze the performance of the proposed time-dependent Poisson-Gamma model in a simulation study considering different types of curves for the underlying recruitment intensity and we compare it to the standard PG model. The model is assessed in terms of percentage bias and coverage rate by the 95\% credible intervals of the observed future recruitments. 

\subsection{The generative model}

\begin{figure}[ht]
  \centering
\includegraphics[width=\linewidth]{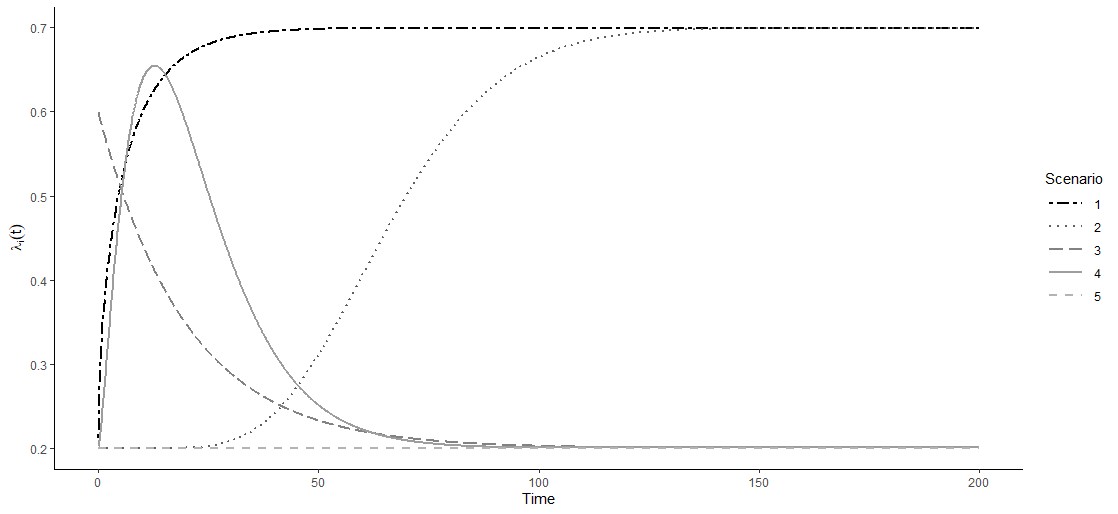}
\caption{Average recruitment curves over time.}
\label{fig:curves}
\end{figure}

The recruited number of participants at each time point $t$ and for each center $i$ is drawn from a Poisson distribution with rate $\widetilde{\lambda}_i(t)$, where $\widetilde{\lambda}_i(t)$ represents a draw from
\[
\lambda_i(t)\sim \textrm{Gamma}\left(\alpha, \frac{\alpha}{f^q(t)}\right),\quad i=1,\dots, C, \,\, t=1,\dots,t_p
\]
and $\widetilde\lambda_i(t)=\widetilde\lambda_i(t_p)$ for $t>t_p$.
The mean function $f^q(t)$ is the shifted scaled probability density function ($q=1$) or the CDF ($q=2$) of a Gamma random variable with various parameter choices, that is
\begin{align*}
f^1(x)&=c_1 + c_2\frac{p_2^{p_1}}{\Gamma(p_1)}e^{-p_2 x}x^{p_1-1},\\
f^2(x)&=c_1 + c_2\frac{p_2^{p_1}}{\Gamma(p_1)}\int_0^{x}e^{-p_2 u}u^{p_1-1}du.
\end{align*}
Varying the parameters of this generative model, we define five recruitment scenarios. The full list of parameters is provided in Table \ref{par} and the resulting average recruitment curves are depicted in Figure \ref{fig:curves}.
\begin{table}[ht] \centering 
  \caption{Simulation parameters.} 
  \label{par} 
\begin{tabular}{@{\extracolsep{5pt}} cccccccc} 
\\[-1.8ex]\hline 
\hline 
 Scenario & $c_1$ & $c_2$ & $p_1$ & $p_2$ &$t_p$ & $q$&$\alpha$ \\ 
 \hline \\[-1.8ex] 
1 &0.2 &0.50 & 0.55 &0.09&60&1&1\\
2 & 0.2 &0.50 & 10 &0.15&150&1&1\\
3 & 0.2 &8 & 1&0.05&130&2&1\\
4&0.2 &13 & 2.40&0.11&100&2&1\\
5 &0.2 &0 & NA &NA&1&NA&1\\
\hline \\[-1.8ex] 
\end{tabular} 
\end{table}
The parameters were selected in order to capture different common recruitment behaviors. The first two scenarios describe a slow start, and the use of the Gamma CDF to capture this evolution over time has been adopted in Deng \& al.~\cite{deng} and Zhang \cite{zhang}. On the other hand, Scenarios 3 and 4 describe a fast start. We have included these enrollment scenarios after private discussions with project coordinators of studies where the recruitment process was aided by modern instruments such as social media. These studies experienced a fast start or an early bump in recruitment that was followed by the attainment of a plateau. Two of the studies where these behaviors were observed, for example, are the CanDirect study \cite{cand} and the Canadian Co-infection Cohort study analyzed in Section 4.
Finally, Scenario 5 describes the constant accrual setting that constitutes the framework of several recruitment models including the standard PG methodology. We evaluate the model for 20 and 60 centers in Setting 1 and 30 and 60 centers in Setting 2. We select two values of $t_{int}$, i.e. $t^1_{int}$ and $t^2_{int}$, whose values depend on the scenario to assess the performance of the model when more information is collected.\\
Additionally, for each scenario, we consider two settings:
\begin{itemize}
    \item Setting 1: all the centers start at the same time, i.e. $u_i=1$ for all $i$.
    \item Setting 2: initiation times are staggered.
\end{itemize}
In Setting 2, we assume that the center-specific initiation times are distributed between $t=1$ and the first interim time $t^1_{int}$, so that there are some centers whose recruitment intensity over time has not reached the plateau point. Two cases are considered: in Case 1, half of the centers have passed the plateau point ($C^*=\frac{1}{2}C$) by the first interim time $t^1_{int}$, whereas in Case 2 this proportion increases to two-thirds ($C^*=\frac{2}{3}C$). To ensure the consistency of these proportions across data replications, the initiation times of the first $C^*$ centers are sampled from a Uniform distribution on $[1,t^1_{int}-t_p]$, and the remaining $C-C^*$ starting times from a Uniform density on $(t^1_{int}-t_p, t^1_{int})$. Therefore, by the time point $t^2_{int}$, some centers switch from the second group to the first one, i.e. $C^*$ increases as $t^1_{int}\rightarrow t^2_{int}$. Given the structure of Setting 2, Scenario 5 is not considered as its plateau point is set at the start of the recruitment phase. In summary, we generated five scenarios in Setting 1 and four in Setting 2. For each combination of scenarios and settings, we consider two interim times, two values of $C$, and two cases for Setting 2.\\
As for the choice of spline models for $m(\pmb{\phi},t)$, we fitted four B-splines models varying the degree of the polynomials and the number of knots. To limit the dimension of the vector of parameters, we considered quadratic and cubic splines and one and zero internal knots, where the placement of the knot was set to $\widehat{t}_p /2$. The best-fitting model among the four candidates was selected using the BIC. The results are based on 1000 data replications.

\subsection{Results}

Table \ref{ci_1}  illustrates the performance of the proposed time-dependent Poisson-Gamma (tPG) model and the standard PG technique in Setting 1 in terms of coverage rate of the observed recruitments by the 95\% credible interval (CrI), percentage bias, and standard error (SE). Table \ref{ci_2}  provides the same information for the setting with staggered initiation times. 
\begin{table}[H] \centering 
  \caption{Coverage rate (CR) of the 95\% CrI, percentage bias, and standard error (SE) over 1000 data replications under  the proposed method (tPG) and the standard Poisson-Gamma (PG) model in Setting 1 (same initiation time across centers).} 
  \label{ci_1} 
\begin{tabular}{@{\extracolsep{5pt}} cccc  cc   cc} 
\\[-1.8ex]\hline 
\hline \\[-1.8ex] 
&&&&\multicolumn{2}{c}{tPG} & \multicolumn{2}{c}{PG}\\
&Scenario&$T$&$t_{int}$&\multicolumn{2}{c}{C} &\multicolumn{2}{c}{C} \\
&&&&20 & 60 &20 & 60  \\
\hline \\[-1.8ex] 
\multirow{10}{*}{CR}&\multirow{ 2}{*}{1}&\multirow{ 2}{*}{300}&80& $0.93$ & $0.95$& $0.27$ & $0.24$\\ 
&&&160&$0.94$ &$0.96$&$0.48$ & $0.47$ \\ 
&\multirow{ 2}{*}{2}&\multirow{ 2}{*}{500}&200&$0.94$ & $0.95$& $0.07$ & $0.01$\\
&&&300&$0.95$ & $0.95$& $0.12$ & $0.01$\\ 
&\multirow{ 2}{*}{3}&\multirow{ 2}{*}{500}&160&$0.90$ & $0.92$& $0.16$ & $0.04$\\ 
&&&250&$0.94$ & $0.95$& $0.28$ & $0.08$\\ 
&\multirow{ 2}{*}{4}&\multirow{ 2}{*}{400}&120&$0.88$ & $0.90$& 0.02 &0\\ 
&&&200&$0.96$ & $0.94$&0.06  & 0 \\ 
&\multirow{ 2}{*}{5}&\multirow{ 2}{*}{300}&80& 0.96 & 0.94& 0.95 &0.95\\ 
&&&160& 0.95 &0.96 &0.94  & 0.93 \\ 
\hline \\[-1.8ex] 
\multirow{10}{*}{Bias}&\multirow{ 2}{*}{1}&\multirow{ 2}{*}{300}&80& $4.98$ & $2.93$& $14.05$ & $8.53$\\ 
&&&160&$2.98$ &$1.57$&$7.30$ & $4.40$  \\ 
&\multirow{ 2}{*}{2}&\multirow{ 2}{*}{500}&200&$3.31$ & $1.89$& $22.65$ & $22.78$\\ 
&&&300&$2.31$ & $1.33$&$15.16$ & $15.28$\\ 
&\multirow{ 2}{*}{3}&\multirow{ 2}{*}{500}&160&$7.92$ & $4.53$& $30.72$ & $26.41$\\ 
&&&250&$4.45$ & $2.53$&$19.89$ & $16.89$\\ 
&\multirow{ 2}{*}{4}&\multirow{ 2}{*}{400}&120&$9.96$ & $5.43$& $62.64$ & $55.76$\\ 
&&&200&$4.77$ & $2.96$& $37.72$ & $33.52$  \\ 
&\multirow{ 2}{*}{5}&\multirow{ 2}{*}{300}&80&5.25 &3.08 & 5.33 &2.99\\ 
&&&160& 4.71 &2.72 &4.88  & 2.76 \\ 
\hline \\[-1.8ex] 
\multirow{10}{*}{SE}&\multirow{ 2}{*}{1}&\multirow{ 2}{*}{300}&80& $660$ & $1187$& $211$ & $375$ \\ 
&&&160&$431$ &$758$&$280$ & $493$  \\ 
&\multirow{ 2}{*}{2}&\multirow{ 2}{*}{500}&200&$907$ & $1591$& $247$ & $441$\\ 
&&&300&$610$ & $1064$& $315$ & $558$\\ 
&\multirow{ 2}{*}{3}&\multirow{ 2}{*}{500}&160&$276$ & $507$& $92$ & $156$ \\ 
&&&250&$220$ & $388$& $115$ & $201$\\ 
&\multirow{ 2}{*}{4}&\multirow{ 2}{*}{400}&120&$230$ & $417$& $86$ & $146$\\ 
&&&200&$179$ & $306$&$97$ & $164$\\ 
&\multirow{ 2}{*}{5}&\multirow{ 2}{*}{300}&80& 204 & 356& $203$ & $353$\\ 
&&&160& 128 &224 &$126$ & $222$ \\ 
\hline \\[-1.8ex] 
\end{tabular} 
\end{table}
In general, the tPG model led to a significant improvement with respect to the PG method in Scenarios 1-4 in terms of CrIs coverage rates and percentage bias. As expected, the standard Poisson-Gamma model failed to deliver acceptable results, as it generated biased estimates in the four scenarios where the recruitment intensity is time-dependent, with the percentage bias nearing or exceeding 50\% and the CrIs coverage rate reaching 0 in several instances. On the other hand, the time-dependent Poisson-Gamma model was able to capture the non-constant evolution of the recruitment intensity over time providing more accurate predictions. 
Additionally, in Scenario 5, the tPG and PG methods led to similar performances, indicating the suitability of the proposed methodology even in the scenario where the assumptions of the PG model are satisfied. 

\begin{table}[H] \centering 
  \caption{Coverage rate (CR) of the 95\% CrI, percentage bias, and standard error (SE)  over 1000 data replications under  the proposed method (tPG) and the standard Poisson-Gamma (PG) model in Setting 2 (staggered initiation times).} 
  \label{ci_2} 
\begin{tabular}{@{\extracolsep{5pt}} cccc  cccc  cccc} 
\\[-1.8ex]\hline 
\hline \\[-1.8ex] 
&&&&\multicolumn{4}{c}{tPG} & \multicolumn{4}{c}{PG}\\
&Scenario&$T$&$t_{int}$&\multicolumn{2}{c}{Case 1} & \multicolumn{2}{c}{Case 2}&\multicolumn{2}{c}{Case 1} & \multicolumn{2}{c}{Case 2}\\
&&&&C=30 & C=60 &C=30 &C=60&C=30 & C=60 &C=30 &C=60  \\
\hline \\[-1.8ex] 
\multirow{ 8}{*}{CR}&\multirow{ 2}{*}{1}&\multirow{ 2}{*}{300}&80&$0.88$ & $0.91$ & $0.91$ & $0.92$ &$0.05$ & $0.02$ & $0.09$ & $0.03$  \\ 
&&&110&$0.93$ & $0.94$ & $0.94$ & $0.93$ & $0.05$ & $0.01$ & $0.07$ & $0.03$ \\ 
&\multirow{ 2}{*}{2}&\multirow{ 2}{*}{500}&200&$0.87$ & $0.85$ & $0.89$ & $0.86$ &$0$ & $0$ & $0$ & $0$\\ 
&&&260&$0.90$ & 0.90 & $0.92$ & 0.92&$0$ & $0$ & $0$ & $0$\\ 
&\multirow{ 2}{*}{3}&\multirow{ 2}{*}{500}&160&$0.73$ & $0.71$ & $0.84$ & $0.83$&$0.20$ & $0.17$ & $0.18$ & $0.14$\\ 
&&&220&$0.89$ & $0.89$ & $0.92$ & $0.92$&$0.31$ & $0.32$ & $0.30$ & $0.31$ \\ 
&\multirow{ 2}{*}{4}&\multirow{ 2}{*}{400}&120&$0.77$ & $0.74$ & $0.80$ & $0.75$&$0.03$ & $0$ & $0.06$ & $0.01$  \\ 
&&&170&$0.87$ & $0.86$ & $0.89$ & $0.86$&$0.10$ & $0.03$ & $0.13$ & $0.06$  \\ 

\hline \\[-1.8ex] 
\multirow{ 8}{*}{Bias}&\multirow{ 2}{*}{1}&\multirow{ 2}{*}{300}&80&$12.12$ & $8.43$ & $10.53$ & $7.15$ &$26.84$ & $27.79$ & $22.44$ & $22.85$\\ 
&&&110&$8.35$ & $5.75$ & $6.92$ & $4.82$&25.04& 25.67 &20.58 &20.88 \\ 
&\multirow{ 2}{*}{2}&\multirow{ 2}{*}{500}&200&$13.44$ & $9.51$ & $10.57$ & $7.74$&$53.88$ & $54.86$ & $46.41$ & $47.49$ \\ 
&&&260&$10.29$ & 7.53 & $8.28$ & 5.79& 43.97& 44.82& 37.99& 38.82 \\ 
&\multirow{ 2}{*}{3}&\multirow{ 2}{*}{500}&160&$15.97$ & $11.93$ & $12.05$ & $8.98$&$22.22$ & $16.75$ & $23.80$ & $19.44$  \\ 
&&&220&$9.39$ & $7.02$ & $7.58$ & $7.60$&13.59 & 9.54& 13.96 & 9.98\\ 
&\multirow{ 2}{*}{4}&\multirow{ 2}{*}{400}&120&$16.14$ & $12.37$ & $14.54$ & $10.91$&$52.13$ & $48.12$ & $47.23$ & $41.80$ \\ 
&&&170&$9.80$ & $7.48$ & $8.50$ & $6.27$&32.00& 28.79& 29.78 &25.78 \\ 

\hline \\[-1.8ex] 
\multirow{ 8}{*}{SE}&\multirow{ 2}{*}{1}&\multirow{ 2}{*}{300}&80&$708$ & $967$ & $731$ & $1003$  &$270$ & $358$ & $217$ & $298$\\ 
&&&110&$679$ & $949$ & $677$ & $958$ &$220$ & $300$ & $242$ & $340$ \\ 
&\multirow{ 2}{*}{2}&\multirow{ 2}{*}{500}&200&$946$ & $1498$ & $1011$ & $1484$&$198$ & $276$ & $196$ & $270$ \\ 
&&&260&$802$ & $1212$ & $855$ & $1212$& $244$ & $367$ & $273$ & $392$ \\ 
&\multirow{ 2}{*}{3}&\multirow{ 2}{*}{500}&160&$311$ & $453$ & $299$ & $429$&$210$ & $282$ & $149$ & $220$  \\ 
&&&220&$291$ & $399$ & $288$ & $417$ &$98$ & $132$ & $99$ & $141$ \\ 
&\multirow{ 2}{*}{4}&\multirow{ 2}{*}{400}&120&$312$ & $514$ & $308$ & $462$&$159$ & $216$ & $193$ & $258$ \\ 
&&&170&$245$ & $385$ & $258$ & $364$&$96$ & $132$ & $93$ & $128$ \\ 
\hline \\[-1.8ex] 
\end{tabular} 
\end{table} 
Overall, Table \ref{ci_1}  shows that the tPG model performed well in every scenario being considered in Setting 1, as it led to adequate levels of percentage bias and coverage rates which improved as the number of centers increments or more information on the recruitment process is accrued by moving forward the interim time $t_{int}$. However, when the initiation times are staggered, the proposed model's performance decreased. Specifically, the model leads to lower levels of coverage rates of the observed recruitments by the credible intervals, and the percentage bias can reach values over 10\%. This decline in forecasting accuracy is more evident in Scenarios 3 and 4 when the proportion of centers whose recruitment intensity has not reached the plateau is the highest, i.e.~in Case 1 for the first interim time. The model's performance greatly improves as the proportion of centers that belong to $C^*$ increases either by switching from Case 1 to Case 2 or by delaying the interim time. 
Finally, it is of interest to look at which models were selected by the BIC. By and large, the BIC showed an overwhelming preference for smaller models with no internal knots in Scenarios 1, 2, and 5, whereas in Scenarios 3 and 4  models that include an internal knot were chosen more frequently.

\section{Forecasting recruitments in the Canadian Co-infection Cohort study}\label{ccc}

\begin{figure}
  \centering
\includegraphics[width=\linewidth]{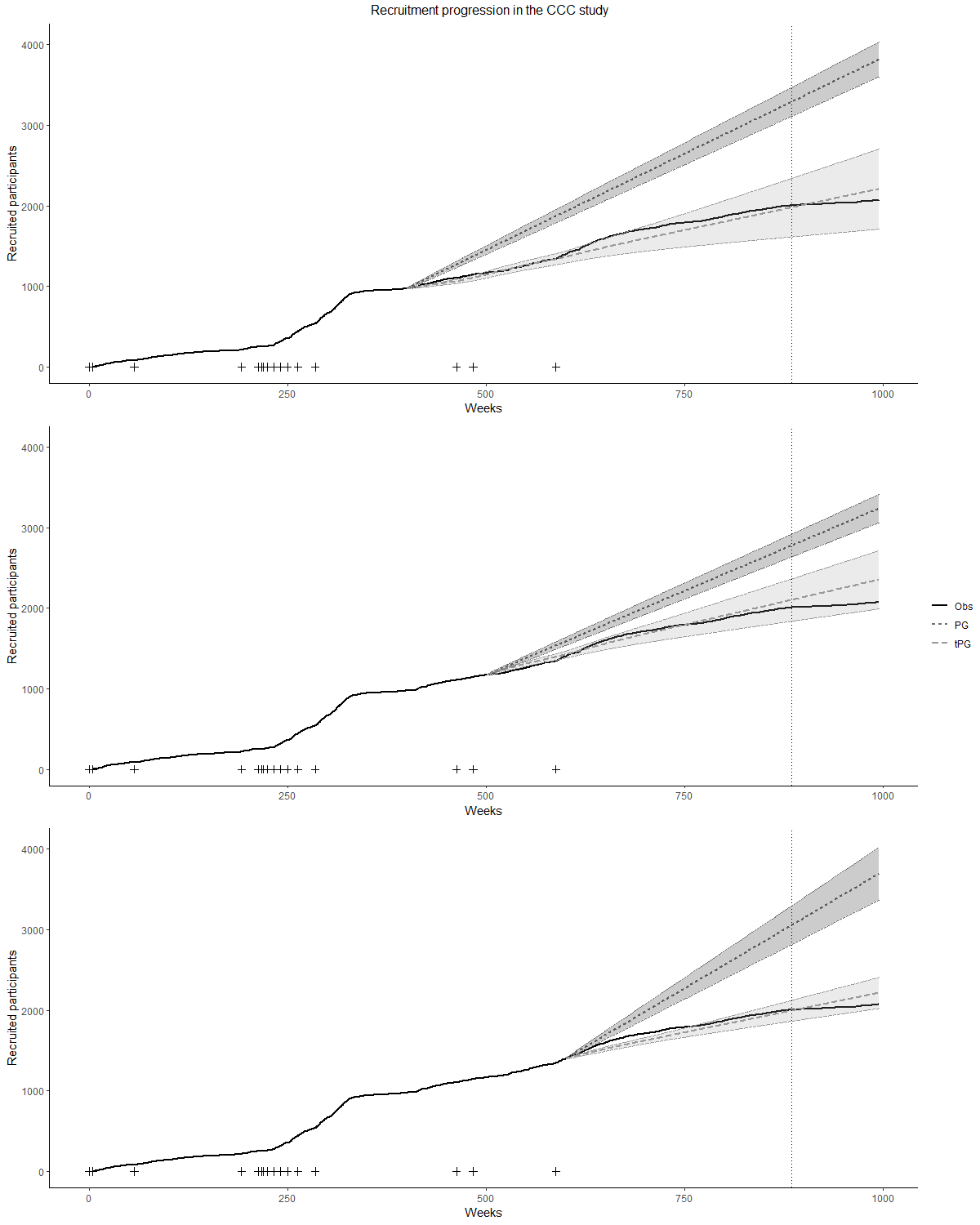}
\caption{Observed and estimated recruitments under the time-dependent Poisson-Gamma (tPG) and Poisson-Gamma (PG) models in the CCC study at three interim times ($t_{int}= 400, 500, \text{and } 600$ weeks). The shaded regions represent the 95\% credible intervals, the `+' signs the opening date for each center, and the dotted vertical line the start of the COVID-19 pandemic.}
\label{fig:ccc}
\end{figure}

In this section, we utilize the time-dependent Poisson-Gamma model to estimate enrollments in the Canadian Co-infection Cohort (CCC) study. The CCC is a prospective observational study whereby participants living with both HIV and Hepatitis C (HCV) are recruited and monitored through follow-up visits scheduled every 6 months.\cite{ccc} The primary goal of the study is to achieve a better understanding of the risk factors associated with liver disease and its progression in the growing population of HCV–HIV co-infected people, with particular emphasis on the effect of highly active antiretroviral therapy (HAART) and HCV treatment. The CCC study encompasses 19\footnote[1]{Two centers were merged but are still treated separately in this analysis.} recruitment centers across Canada opened between March 2003 and June 2014 and, as of April 8th, 2022, it comprises a total of 2077 recruited participants in 995 weeks. \\
To assess the applicability and accuracy of the proposed methodology in forecasting the enrollments in this study, we considered three interim times $t_{int}$. At each of these time points, the accrued data are used to estimate the recruitment progression up to April 8th, 2022, i.e. week 995, and the resulting forecasts are compared to the observed enrollments. Since 13 of the 19 centers were opened between week 192 (November 2006) and 285 (September 2008), we set the interim times to $t_{int}= 400, 500, \text{and } 600$ weeks. The results under the proposed methodology and the standard Poisson-Gamma model are illustrated in Figure \ref{fig:ccc}.
As we can see from the observed recruitment progression, the constant-rate assumption is not met. Specifically, the centers generally showed a higher recruitment intensity in the initial stages of their enrollment period, followed by a deceleration until the attainment of a plateau, which our model consistently estimated at 110 weeks after the centers' opening for every interim time. The violation of the constant-rate assumption is especially visible after the opening of the cluster of centers between weeks 192 and 285, as the recruitment progression visibly decelerates around week 330. Regardless of the interim time, the tPG model led to an estimated recruitment progression that closely tracks the observed enrollments and, apart from brief time periods immediately following the interim times, the resulting 95\% credible intervals covered the observed enrollments. Note that the last two years of recruitment were affected by the COVID-19 pandemic, whose start is identified in Figure \ref{fig:ccc} as March 11th, 2020, i.e. the day the WHO declared the COVID-19 outbreak a global pandemic. The pandemic caused a deceleration in enrollments, and its beginning coincides with the period where the point estimates produced by the tPG model start to deviate the most from the observed recruitments, especially for $t_{int}=500$.\\
On the other hand, the forecasts under the standard Poisson-Gamma model are heavily biased. More specifically, since the centers showed an initial fast recruitment intensity followed by a slowdown, the standard PG model led to an overestimation of the recruitment rates and hence to overly-optimistic forecasts.

\section{Discussion}

In this paper, we have introduced an extension to the Poisson-Gamma recruitment model where the recruitment intensity is allowed to vary over time. Specifically, with respect to the standard methodology where recruitment rates are assumed to be constant, we introduced the presence of a time window that spans from the centers' initiation to an unknown time point where the recruitment intensity is time-dependent. In order to guarantee the applicability of this methodology for a wide range of recruitment scenarios, we modeled the mean parameter of the common Gamma distribution which generates the recruitment rates through B-splines. We analyzed the performance of the proposed methodology in a simulation study for various behaviors of the recruitment curves. The model performed well in every scenario being considered when the centers share the same initiation time. On the other hand, the setting where the centers' initiation times are staggered showed a decline in performance for some scenarios. Specifically, the proposed model struggled to deliver overall satisfactory results when a sizable number of centers has not reached the plateau point in their enrollment process, especially for the more elaborate recruitment behaviors assumed in Scenarios 3 and 4. The model's performance greatly improved when the proportion of these centers was reduced or the interim time was delayed (hence still decreasing this proportion), which suggests that while the model correctly estimates the constant section of the recruitment curves, it may not adequately catch the non-constant section in some scenarios. Note that due to the computational time and the variety of settings considered, we limited the number of spline models to four. Adding more models to the list of candidates by varying the number of internal knots or/and their location might improve the model's accuracy. This is partially confirmed by the BIC model choice. In fact, the BIC selected mostly smaller models to estimate the simpler recruitment curves, whereas larger models with an internal knot were preferred more frequently in the scenarios characterized by more intricate recruitment progressions. However, even in the scenarios where the model's performance could be improved, the proposed recruitment model still led to significantly better results than the standard Poisson-Gamma model. Additionally, in the scenario where the constant-intensity assumption is met, the time-dependent Poisson-Gamma model led to comparable results, suggesting that the proposed methodology can be seen as a time-dependent generalization of the standard PG model without any evident loss in performance.\\
The application of the tPG model to the recruitment data of the Canadian Co-infection Cohort study showcased its suitability and accuracy in forecasting enrollments in a real study even for a low number of recruitment centers (16 for the first interim time), while also highlighting the limitations of the standard PG model when the constant-rate assumption is not met.\\
Finally, some other limitations of the methodology we have presented in this paper need to be highlighted. In particular, this model relies on the availability of recruitment data past the plateau point for an adequate number of centers, which makes it impractical to employ in the planning phases of a clinical study or during its early monitoring stage. The integration of prior information from similar studies or the centers' history is an interesting topic for further research.
Moreover, we assumed that the plateau point is the same for all centers. The addition of a layer of variability to reflect the variations across centers' could represent a relevant improvement in the flexibility of this methodology.

\section*{Acknowledgements}
AT is supported by a doctoral fellowship from the Fonds de Recherche du Québec (FRQ) - Nature et Technologie and by funds from a Recherche en Équipe grant from the FRQ – Nature et Technologie awarded to DAS. DAS and EEMM acknowledge support from Discovery Grants from the Natural Sciences and Engineering Research Council of Canada (NSERC). EEMM is a Canada Research Chair (Tier 1) in Statistical Methods for Precision Medicine and acknowledges the support of a chercheur de mérite career award from the FRQ - Santé.
MBK is supported by a Tier I Canada Research Chair in Clinical and Epidemiologic Studies of Chronic Viral Infections in Vulnerable Populations. The Canadian Coinfection Cohort is funded by the Réseau sida/maladies infectieuses du Fonds de Recherche du Québec—Santé; the Canadian Institute for Health Research (CIHR; FDN-143270); and the CIHR Canadian HIV Trials Network (CTN222).

\section*{Data availability statement}
The data that support the findings in this paper are not shared as restrictions apply to the availability of these data, which were used under license in this paper.

\bibliographystyle{unsrt}
\bibliography{bibliography}

\end{document}